\documentclass[
reprint,
superscriptaddress,
amsmath,
amssymb,
prb,
]{revtex4-1}
\usepackage{graphicx}
\usepackage{dcolumn}
\usepackage{bm}
\usepackage{hyperref}

\begin{document}

\title{First-order correction to the Casimir force within an inhomogeneous medium}
\author{Fanglin Bao}
\affiliation{Department of Physics, Zhejiang University, Hangzhou 310058, China}
\affiliation{Centre for Optical and Electromagnetic Research, JORCEP, Zhejiang University (ZJU), Hangzhou 310058, China}
\author{Bin Luo}
\affiliation{Centre for Optical and Electromagnetic Research, ZJU-SCNU Joint Research Center of Photonics, South China Normal University, Guangzhou 510000, China}
\author{Sailing He}
\affiliation{Centre for Optical and Electromagnetic Research, JORCEP, Zhejiang University (ZJU), Hangzhou 310058, China}
\affiliation{Centre for Optical and Electromagnetic Research, ZJU-SCNU Joint Research Center of Photonics, South China Normal University, Guangzhou 510000, China}
\affiliation{Department of Electromagnetic Engineering, Royal Institute of Technology, 10044 Stockholm, Sweden}

\begin{abstract}\noindent
For the Casimir piston filled with an inhomogeneous medium, the Casimir energy is regularized and expressed with cylinder kernel coefficients by using the first-order perturbation theory. When the refraction index of the medium is smoothly inhomogeneous (i.e., derivatives of all orders exist), logarithmically cutoff-dependent term in Casimir energy is found. We show that in the piston model this term vanishes in the force and thus the Casimir force is always cutoff-independent, but this term will remain in the force in the half-space model and must be removed by additional regularization. We investigate the inhomogeneity of an exponentially decaying profile, and give the first-order corrections to both free Casimir energy and Casimir force. The present method can be extended to other inhomogeneous profiles. Our results should be useful for future relevant calculations and experimental studies.
\end{abstract}

\maketitle

\section{\label{sec:introduction}introduction}
Casimir effect \cite{Casimir1948,Milton2001} is known as one of the direct manifestation of the vacuum zero-point energy in quantum physics. A mode-summation method can be used to predict easily an attractive force between two electrically neutral, perfectly conductive half-spaces. This ghostly force arises from the quantum fluctuations, and thus is ubiquitous in the physical world. However, once predicted, the Casimir force in a variety of scenarios has witnessed many divergence problems during calculations, due to the geometry of boundaries or topology of space \cite{Deutsch1979,Bender1994,Milton2004a}. These unresolved divergences usually show a logarithmic cutoff dependence and seem to be irremovable (while other quartic or cubic cutoff-dependent terms in Casimir energy are already well understood as volume energy or surface energy, and thus are removable) \cite{Fulling2003,Bordag2009}.

When the inhomogeneity of the medium rather than complicated geometry of objects in Casimir apparatus is considered, it is known that analytical description of the Casimir force (stress or force density) has already been obtained \cite{Philbin2010,Rosa2011}, without explicit divergence problem. The results in both references have subtracted the ``bulk contribution" and thus relate only to the scattering Green's function. This, in our minds, regularizes the quartic diverging term (volume energy) in Casimir energy, but might not be a thorough regularization, as also mentioned in Ref.~\cite{Leonhardt2007,Philbin2010}. In fact, divergences indeed occur when one tries to evaluate the numeric values of both Casimir energy and Casimir force, according to the obtained analytical description within an inhomogeneous medium \cite{Philbin2010}. Efforts have been made, since then, to understand the remaining divergence \cite{Simpson2013}, but the problem is still far from solved.

We note, in the model of half-spaces which is used in Ref.~\cite{Philbin2010}, cutoff-dependent terms in Casimir energy must be assigned physical meanings and thus removed manually by introducing corresponding additional regularizations. While in the model of Casimir piston, cutoff-dependent terms may vanish automatically in the Casimir force, due to the cancellation of contributions from the left and right cavities. We also note, a naive mode-summation approach \cite{Horsley2013} in the first-order perturbation theory (which is supposed to show the cutoff dependence), turns out to yield cutoff-independent result of Casimir force for inhomogeneous media. All of the above have led us to expect that more useful information could be obtained from the Casimir piston model. Therefore, we adopt the piston model here, following the mode-summation approach in the first-order perturbation theory as Ref.~\cite{Horsley2013}(but in a more general form), to investigate the Casimir physics within inhomogeneous media. The purposes for doing so are threefold. First, we want to analyze the inhomogeneity-induced Casimir divergence with the heat kernel expansion, and expect to obtain some insights for additional regularizations for half-space model. Secondly, we want to show that the cutoff independence of Casimir force is true for various inhomogeneous profiles, instead of a particular case. Thirdly, we want to show how large the influence of the inhomogeneity on the total Casimir force is. As weak force measurements have been developed and reached a quite precise level (within $1\%$) \cite{Mohideen1998}, and Casimir force between bodies in a liquid has also been measured \cite{Munday2007,Munday2009}, inhomogeneity-induced corrections may be useful for future experimental studies as the experimental configuration becomes more and more complicated. 

In the present paper, we first derive general expressions of Casimir energy for inhomogeneous media in Sec.~\ref{sec:CEIM}. The summations in expressions of Casimir energy are organized and re-expanded over the cutoff parameter in Sec.~\ref{sec:SE}. Then we prove the cutoff independence of Casimir force for smooth inhomogeneity in Sec.~\ref{sec:CI}, where we also see logarithmic divergence in the half-space model. We investigate the exponentially decaying inhomogeneity in Sec.~\ref{sec:application} and give first-order corrections to Casimir energy and Casimir force. Discussions and conclusions are given in Sec.~\ref{sec:discussion} and Sec.~\ref{sec:conclusion}, respectively.
\section{\label{sec:CEIM}Casimir energy of plates within inhomogeneous media}
\begin{figure}[t]
\scalebox{1}{\includegraphics{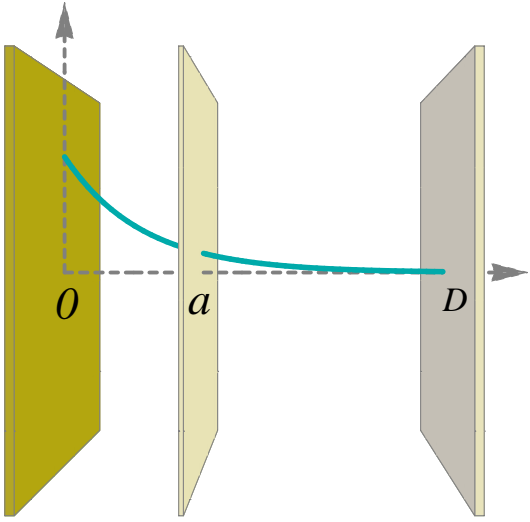}}
\caption{\label{fig:piston}Casimir piston model \cite{Horsley2013} with four (up, bottom, front and back) hidden plates. The cyan curve represents an arbitrary inhomogeneous profile. All plates are of perfect conductivity.}
\end{figure}

In the mode-summation technique, total zero-point energy of the Casimir piston device, see, Fig.~\ref{fig:piston}
, is expressed as \cite{Horsley2013}
\begin{equation}\label{eq:zpe}
E_0=\frac12\sum\limits_{m,p,q,\lambda}{\omega_{m,p,p,\lambda}+\mathcal{L}\to\mathcal{R},}
\end{equation}
where $m,p,q$ are indexes for three wave numbers ($m$ for the direction perpendicular to the plates), $\lambda$ is the index for polarization, notation $\mathcal{L}\to\mathcal{R}$ represents the counterpart for the right cavity, and also we have set $\hbar=c=1$ which will be recovered later according to dimensional analysis. For simplicity, we use $k_\parallel$ to account for $\{p,q\}$, $t$ for $\{m,k_\parallel\}$ and $J$ for $\{t,\lambda\}$. Below we will focus on the left cavity and omit the notation $\mathcal{L}\to\mathcal{R}$ in all equations, just keeping in mind that the right counterpart should be added in the final step. We adopt the cutoff regularization, Eq.~\ref{eq:zpe} then becomes
\begin{equation}\label{eq:re}
\tilde{E}=\lim\limits_{\xi\to0}{\frac12\sum\limits_J{\omega_J e^{-\xi\omega_J}}}.
\end{equation}
We omit the limit notation in the following for simplicity as well.

When there is an inhomogeneous perturbation in refraction index $n(x)=n_0[1+\delta\alpha\cdot f(x)]$, where $\delta\alpha$ is a small perturbation value, the difference of regularized Casimir energy is then
\begin{equation}\label{eq:dre}
\delta\tilde{E}=\frac12\partial_\xi\sum\limits_J{\xi\omega^1_J e^{-\xi\omega^0_J}},
\end{equation}
with $\omega^1_J=-\delta\alpha\mathcal{P}_J\omega^0_J$ and $\mathcal{P}_J \equiv \langle\chi^0_J|f(x)|\chi^0_J\rangle$. $\chi^0_J$ is the $J_{th}$ unperturbed eigen-wavefunction with $\omega^0_J$ being its eigen-frequency. We can find the expressions for the electric fields $\chi^0_J$ in Ref.~\cite{Horsley2013} (Eqs.~(14) and (15) after correction in their erratum; we do not use $E$ for electric field here to avoid confusion with Casimir energy). We have also considered the perturbation of frequency in the exponent, and thus we get a factor of $\partial_\xi\cdot\xi\cdot$ in our Eq.~\ref{eq:dre}. However it does not otherwise change our argument. Therefore, we can obtain
\begin{equation}\label{eq:dre2}
\delta \tilde{E}=-\frac{\delta\alpha}{2n_0}\partial_\xi\cdot\xi \sum\limits_J{\mathcal{P}_J k^0_J e^{-\xi k_J^0/n_0}}.
\end{equation}
Here $k_J^0=\sqrt{k^2_\parallel+(m\pi /a)^2}$ is the wave number for homogeneous media defined as $k^0_J \equiv n_0\omega^0_J /c$, and we have set $c=1$. The perturbation theory is justified as long as $\mathcal{P}_J$ is bounded and $\delta\alpha\mathcal{P}_J\ll1$. We note that only $\mathcal{P}_J$ varies for different polarizations. To sum up polarizations first, we obtain
\begin{equation}\label{eq:polarization}
\sum_{\lambda=1,2}{\mathcal{P}_J} = (2-\delta_{m0})F_0 - (2-\delta_{m0})\frac{(m\pi/a)^2}{k_\parallel^2+(m\pi/a)^2}F_m,
\end{equation}
with the Fourier coefficient of perturbation profile $f(x)$
\begin{equation}\label{eq:fourier}
F_m \equiv \frac 1a \int_0^a{f(x) \cos{\frac{2m\pi x}a} \rm{d}x},
\end{equation}
where $\lambda=1,2$ represents two different polarizations. To obtain the counterparts of our Eqs.~\ref{eq:polarization} and \ref{eq:fourier} for the right cavity we can substitute $a$ with $D-a$ and also change the integration range in Eq.~\ref{eq:fourier} from $(0,a)$ to $(a,D)$.

Substituting Eq.~\ref{eq:polarization} into Eq.~\ref{eq:dre2}, we can split the Casimir energy into two parts for the convenience of calculation, the first part is
\begin{equation}\label{eq:dreI}
\begin{split}
\delta \tilde{E}_1 &= -\frac{\delta\alpha}{2n_0} \partial_{\xi}\xi \sum\limits_t{
(2-\delta_{m0})F_0
 k^0_t e^{-\xi k_t^0/n_0}}\\
&= \frac{\delta\alpha An_0^2}{4\pi} F_0 \partial_{\xi}\xi\partial_{\xi}\hat{\Xi}(\xi) \sum_m{ \left(2-\delta_{m0}\right)e^{-m\pi\xi/an_0}},
\end{split}
\end{equation}
and the second part is
\begin{equation}\label{eq:dreII}
\begin{split}
\delta \tilde{E}_2 &= \frac{\delta\alpha}{2n_0}\partial_{\xi}\xi \sum\limits_t{
(2-\delta_{m0})\frac{(m\pi/a)^2}{k_\parallel^2+(m\pi/a)^2}F_m k^0_t e^{-\xi k_t^0/n_0}}\\
&= \frac{\delta\alpha An_0^2}{4\pi}\partial_{\xi}^3 \sum_m{(2-\delta_{m0})F_m e^{-m\pi\xi/an_0}},
\end{split}
\end{equation}
where $\hat{\Xi}\equiv\xi^{-2}(1-\xi\partial_\xi), m=0,1,2,\cdots$, and $A$ is the surface area of plates. To get the second equalities for both Eqs.~\ref{eq:dreI} and \ref{eq:dreII} we have integrated over $k_\parallel$, provided that the lateral dimension of the plates is much larger than the distance between plates, i.e., $\sqrt{A}\gg D$, so we can replace summation with the corresponding integral. The counterparts of Eqs.~\ref{eq:dreI} and \ref{eq:dreII} for the right cavity can be obtained by substituting $a$ with $D-a$.

On the other hand, following Eq.~\ref{eq:re} and integrating over $k_\parallel$, we can also obtain the regularized Casimir energy for the homogeneous case
\begin{equation}\label{eq:reh}
\tilde{E}_h=-\frac{An_0^2}{4\pi}\partial_\xi \hat{\Xi}(\xi) \sum_m{(2-\delta_{m0})e^{-m\pi\xi/an_0}}.
\end{equation}

Now, we can multiply Eqs.~\ref{eq:dreI}, \ref{eq:dreII} and \ref{eq:reh} with a factor of $\hbar c$, and add to them with the counterparts of the right cavity, and then take the $\xi\to0$ limit. The sum of these three equations gives the total Casimir energy for inhomogeneous case. Its derivative with respect to position $a$ yields the Casimir force on the central plate. These three expressions are quite general for perturbation profiles as long as $\mathcal{P}_J$ is bounded and $\delta\alpha\mathcal{P}_J\ll1$.
\section{\label{sec:SE}summation and re-expansion}
The summations in Eqs.~\ref{eq:dreI} and \ref{eq:reh} can be expressed in terms of polylogarithm function
\begin{equation}\label{eq:suml}
\begin{split}
\sum_{m=0}{(2-\delta_{m0})e^{-m\pi\xi/an_0}}
& =\left(\sum_{m=1}{2e^{-m\pi\xi/an_0}}\right)+(2-\delta_{00})\\
& =2Li_0\left(e^{-\pi\xi/an_0}\right)+1.
\end{split}
\end{equation}
The summation in Eq.~\ref{eq:dreII} depends on the particular form of the perturbation profile. However the $m=0$ term vanishes obviously. Then we can rewrite it as
\begin{equation}\label{eq:dreII2}
\delta\tilde{E}_2=\frac{\delta\alpha An_0^2}{2\pi}\partial_\xi^3 \sum\limits_{m=0 or 1}{F_me^{-m\pi\xi/an_0}},
\end{equation}
where $m$ can run from $0$ or $1$ as needed. For the Taylor basis of order $d$
\begin{equation}\label{eq:taylor}
f_d(x)=\left(\frac xD\right)^d,
\end{equation}
where $d=0,1,2,\cdots$, according to the definition of $F_m$ (we now consider $m$ running from 1), we can integrate by parts to get the following relation for $d\ge 2$ for the left cavity
\begin{equation}
F_{m,d}^\mathcal{L}=\frac{ad}{D(2m\pi)^2}(\frac aD)^{d-1}-(\frac a{2m\pi})^2 \frac{d(d-1)}{D^2}F_{m,d-2}^\mathcal{L},
\end{equation}
while we have $F_{m,0}^\mathcal{L}=F_{m,1}^\mathcal{L}=0$. The explicit expression is then
\begin{equation}\label{eq:fouriertaylorl}
F_{m,d}^\mathcal{L}=\sum\limits_{i=1}^{[d/2]}{\frac{(-1)^{i-1}d!}{(d+1-2i)!}(2m\pi)^{-2i}\left(\frac aD\right)^{2i-1}\left(\frac aD\right)^{d+1-2i}},
\end{equation}
where $[d/2]$ denotes the bare integer part. The counterpart for the right cavity can be obtained similarly
\begin{equation}\label{eq:fouriertaylorr}
\begin{split}
F_{m,d}^\mathcal{R}
&=\sum\limits_{i=1}^{[d/2]}{\frac{(-1)^{i-1}d!}{(d+1-2i)!}(2m\pi)^{-2i}}\\
&\cdot\left(\frac {D-a}D\right)^{2i-1}\left[1-(\frac aD)^{d+1-2i}\right].
\end{split}
\end{equation}
We note that they all have the form $\sum_{i=1}^{[d/2]}{g_im^{-2i}}$, where $g_i$'s are some coefficients independent of $m$. Back to Eq.~\ref{eq:dreII2}, we now know the summation is a combination of polylogarithm functions
\begin{equation}\label{eq:dreII3}
\delta\tilde{E}_2=\frac{\delta\alpha An_0^2}{2\pi}\partial_\xi^3 \sum\limits_{i=1}^{[d/2]}{g_i\cdot Li_{2i}\left( e^{-\pi\xi/an_0}\right)}.
\end{equation}
In general, these polylogarithm functions can be expanded over $\xi$ (with $\xi\to0$) as
\begin{gather}
Li_0\left(e^{-\frac{\pi\xi}{an_0}}\right)=\frac{an_0}{\pi\xi}+\sum\limits_{k=0}^\infty{\frac{\zeta(-k)}{k!}\left(-\frac{\pi\xi}{an_0}\right)^k}, \nonumber \\
\begin{split}\label{eq:expandlog}
Li_s\left(e^{-\frac{\pi\xi}{an_0}}\right)
&=\frac 1{(s-1)!}\left(-\frac{\pi\xi}{an_0}\right)^{s-1}\left[H_{s-1}-\log{(\frac{\pi\xi}{an_0})}\right]\\
&+\sum\limits_{k=0,k\ne s-1}^\infty{\frac{\zeta(s-k)}{k!}\left(-\frac{\pi\xi}{an_0}\right)^k},
\end{split}
\end{gather}
where $s=2,4,6,\cdots$, $H_s=\sum_{h=1}^s{1/h}$ is the harmonic number with $H_0=0$.

Before going further to get the explicit expression of Casimir energy for inhomogeneous case, we inspect another set of profiles
\begin{equation}\label{eq:exponent}
f(x)=e^{-\eta x+\Delta},
\end{equation}
where $Re[\eta]>0$. For the left cavity we have (we now consider $m$ running from 0)
\begin{equation}\label{eq:fourierexponent}
F_{m,\eta}^\mathcal{L}=\frac{e^\Delta}{4\pi i}[1-e^{-\eta a}]\left[\frac 1{m-i\eta a/2\pi}-\frac 1{m+i\eta a/2\pi}\right].
\end{equation}
The right counterpart could be obtained by substitution $(1-e^{-\eta a})\to(e^{-\eta a}-e^{-\eta D}),a\to(D-a)$. Eq.~\ref{eq:dreII2} then becomes
\begin{equation}\label{eq:dreII4}
\begin{split}
\delta\tilde{E}_2
&= \frac{\delta\alpha An_0^2}{2\pi}\frac{e^\Delta}{4\pi i}[1-e^{-\eta a}]\\
&\cdot \partial_\xi^3 \left[\phi(e^{-\frac{\pi\xi}{an_0}},1,-\frac{i\eta a}{2\pi})-\phi(e^{-\frac{\pi\xi}{an_0}},1,\frac{i\eta a}{2\pi}) \right].
\end{split}
\end{equation}
Here Lerch zeta function $\phi$ can be expanded over $\xi$ (with $\xi\to0$) as
\begin{equation}
\begin{split}
\phi(e^{-\frac{\pi\xi}{an_0}},1,\beta)
&= \left[\sum_{k=0}{\frac{(\frac{\pi\xi}{an_0}\beta)^k}{k!}}\right]\Bigg\{\sum_{k=1}{\zeta(1-k,\beta)\frac{(-\frac{\pi\xi}{an_0})^k}{k!}}\\
&+\left[\psi(1)-\psi(\beta)-\log{(\frac \pi{an_0})}-\log\xi \right]\Bigg\},
\end{split}
\end{equation}
where $\psi$ is the digamma function and $\zeta$ is the Hurwitz zeta function. We note $\beta$ is not any negative integer and thus $\zeta(1-k,\beta),\psi(\beta)$ are always finite. Furthermore, since only $\beta$ is complex, we have $\phi(e^{-\pi\xi/an_0},1,\beta^*)=\phi^*(e^{-\pi\xi/an_0},1,\beta)$ and that is why Eq.~\ref{eq:dreII4} is always real.
\section{\label{sec:CI}cutoff independence}
Up to now, we see the total Casimir energy for inhomogeneous case is generally expressed as a Laurent-type-like series
\begin{equation}\label{eq:HKE}
\tilde{E}=\tilde{E}_h+\delta\tilde{E}_1+\delta\tilde{E}_2=\sum_{i=-4}{C_i\xi^i}+\sum_{i=0}{Cl_i\xi^i\log\xi}.
\end{equation}
These coefficients $C_i$ and $Cl_i$ are well-studied heat kernel (or more precisely cylinder kernel here) coefficients in cutoff regularization \cite{Vassilevich2003,Fulling2003,Bordag2009}. They depend only on the geometry property and the boundary condition of the system under consideration. The divergent terms under limit $\xi\to0$ are usually assigned to the self-energy of volume or surface and so on to renormalize the theory. Here, in our case, we should check these divergent terms, making sure that their coefficients are independent of position $a$ so that these divergences would not go into the Casimir force, as we expect the observable---the Casimir force---to be finite. The constant term $C_0$ is the free Casimir energy, and its derivative with respect to position $a$ is the Casimir force we want to calculate. Once these divergent terms are assured to be independent of position $a$, the final result of Casimir force should be identical to the result from Ref.~\cite{Goto2012}.

We now prove the cutoff-independence of Casimir force for any smoothly inhomogeneous perturbation profile. We know, any smooth function in domain $(0,D)$ can be expanded via the Taylor bases given in Eq.~\ref{eq:taylor}. Therefore, equivalently what we need is to prove the cutoff-independence of Casimir force for basis of any order $d$. We proceed in the following way.

First, the cutoff property of Eq.~\ref{eq:dreI} is determined by $F_0\partial_\xi\hat{\Xi}(\xi)(2Li_0+1)$ and the cutoff property of Eq.~\ref{eq:reh} is determined by $\partial_\xi\hat{\Xi}(\xi)(2Li_0+1)$ according to Eq.~\ref{eq:suml}. We note the operators acting on the polylogarithm function generally decrease the power of $\xi$ by three orders, and $Li_0=\sum_{i=-1}{l_i\xi^i}$. Thus we have to check $i=-1$ to $3$ to see the cutoff property. According to the expansion of polylogarithm function, Eq.~\ref{eq:expandlog}, we have
\[
l_{-1}^\mathcal{L}\xi^{-1}=\frac{an_0}{\pi\xi},\  l_{-1}^\mathcal{R}\xi^{-1}=\frac{(D-a)n_0}{\pi\xi}.
\]
Thus for Eq.~\ref{eq:reh} we have
\[
C_{-4}\xi^{-4}\propto ADn^3_0\xi^{-4},
\]
and for Eq.~\ref{eq:dreI} we have
\[
C_{-4}\xi^{-4}\propto ADn^3_0\xi^{-4}\cdot\delta\alpha\frac 1D\int_0^D{f\rm{d}x}.
\]
These two quartic divergent terms serve as the self-energy of the inter-media which is exactly of volume $AD$, and their independence of position $a$ indicates they will not come into the Casimir force. Next, we also have
\[
l_0^\mathcal{L}=l_0^\mathcal{R}=-\frac 12,
\]
which means $C_{-3}\xi^{-3}=0$ for both Eqs.~\ref{eq:dreI} and \ref{eq:reh}. This term is usually proportional to the surface area $A$ and serves as the surface energy. The absence of surface divergence term is due to the cancellation of TE and TM contributions as also reported in Ref.~\cite{Milton2004a}. Next, we note, whatever $l_1\xi^1$ is, we have $\hat{\Xi}(\xi)\cdot l_1\xi^1=0$ and whatever $l_2\xi^2$ is, we have $\partial_\xi\hat{\Xi}(\xi)\cdot l_2\xi^2=0$. Thus $C_{-2}\xi^{-2}=C_{-1}\xi^{-1}=0$. Therefore, the cutoff-dependent terms in Eqs.~\ref{eq:dreI} and \ref{eq:reh} all vanish in the Casimir force.

Now we turn to the contribution from Eq.~\ref{eq:dreII}, which has been expressed as Eq.~\ref{eq:dreII3}. Since $\partial_\xi^3$ has a good property---all polynomial terms under it vanish when $\xi\to0$, for positive plural $s=2i$, we only need to consider logarithm terms in $Li_s$ to check the cutoff property. We have
\begin{equation}
\begin{split}
&\ \ \ \ g_i^\mathcal{L}\left(\frac \xi a\right)^{2i-1}\log\xi+g_i^\mathcal{R}\left(\frac \xi{D-a}\right)^{2i-1}\log\xi\\
& \propto \log\xi\Bigg\{
\left(\frac \xi a\right)^{2i-1} \left(\frac aD\right)^{2i-1} \left(\frac aD\right)^{d+1-2i}\\
& +\left(\frac \xi{D-a}\right)^{2i-1} \left(\frac{D-a}D\right)^{2i-1}\left[1-\left(\frac aD\right)^{d+1-2i}\right]
\Bigg\}\\
& =\left(\frac\xi D\right)^{2i-1}\log\xi. \label{eq:log}
\end{split}
\end{equation}
After performing $\partial_\xi^3$, we see, term $i=1$ contributes to $C_{-2}\xi^{-2}$ and term $i=2$ contributes to $Cl_0\log\xi$. All other logarithm terms under $\partial_\xi^3$ vanish when $\xi\to0$. We see $C_{-2}\propto\frac 1D, Cl_0\propto\frac 1{D^3}$, and thus both of them are independent of position $a$. However, in half-space model there is no right cavity, those logarithm terms will appear to be $a$-dependent and must be removed manually. Unfortunately, this is an unresolved problem yet, as we kown.

This completes our proof that for Taylor basis of any order $d$, the Casimir force for plates within an inhomogeneous medium is cutoff-independent. Therefore, for any smooth inhomogeneity, which is a superposition of Taylor bases, the Casimir force will have the inherited cutoff independence.
\section{\label{sec:application}applications}
The Taylor expansion of inhomogeneity profile, is useful for cutoff analysis, but will result in a series of constant terms $C_0^d$. This is inconvenient for calculation of the free Casimir energy and the Casimir force. Fortunately, for some cases, we are able to do the calculation without the Taylor expansion. One example is the profile given in Eq.~\ref{eq:exponent} with $\eta>0,\Delta=0$.

The cutoff independence of $\tilde{E}_h,\delta\tilde{E}_1$ can be analyzed exactly in the same way as above, while the $\delta\tilde{E}_2$ now is described by Eq.~\ref{eq:dreII4}. Similarly, the logarithm terms
\[
\begin{split}
&(1-e^{-\eta a})\left(\frac{\beta^\mathcal{L}\xi}a \right)^k\log\xi+
(e^{-\eta a}-e^{-\eta D})\left(\frac{\beta^\mathcal{R}\xi}{D-a} \right)^k\log\xi\\
&\propto (1-e^{-\eta D})\xi^k\log\xi,
\end{split}
\]
is independent of position $a$. Therefore, we have again the cutoff independence of Casimir force, as expected.

To obtain the explicit expression of the Casimir force, we only need to calculate the $C_0$ term, which comes from $\xi^3$ in $Li$ or $\phi$. Now we have
\begin{equation}
F_0^\mathcal{L}=\frac 1{\eta a}(1-e^{-\eta a}),\ 
F_0^\mathcal{R}=\frac 1{\eta(D-a)}(e^{-\eta a}-e^{-\eta D}).
\end{equation}
$C_0$ contribution from $\tilde{E}_h$ is
\begin{equation}
\begin{split}
&-\frac{An_0^2}{4\pi}(-2)\left[\frac{2\zeta(-3)}{3!}(-\frac\pi{an_0})^3+\frac{2\zeta(-3)}{3!}(-\frac\pi{(D-a)n_0})^3 \right]\\
&=-\frac{A\hbar c \pi^2}{720n_0}\left[\frac 1{a^3}+\frac 1{(D-a)^3} \right].
\end{split}
\end{equation}
Here we have recovered $\hbar c$ in the end. $C_0$ contribution from $\delta\tilde{E}_1$ is
\begin{equation}
\delta\alpha\frac{A\hbar c\pi^2}{720n_0}\left[F_0^\mathcal{L}(\frac 1a)^3+F_0^\mathcal{R}(\frac 1{D-a})^3\right].
\end{equation}
During the calculation of this term, we have seen the operator $\partial_\xi\cdot\xi\cdot$ introduced by including the exponent in the regularization in Eq.~\ref{eq:dre}, does not change the observable value (compared with the one without including the exponent in the regularization), as expected. $C_0$ contribution from $\delta\tilde{E}_2$ is
\begin{equation}\label{eq:dreIIex}
\begin{split}
&\delta\alpha\frac{A\hbar c\pi^2}{720n_0}(\frac 1a)^3\cdot\frac{180}\pi(1-e^{-\eta a})\cdot\\
&\Im\left[\beta^3\left(\psi(1)-\log{\frac \eta{2n_0}}+\frac{11}6\right)+\beta^3(\log{i\beta}-\psi(\beta))-\beta/12\right]\\
&+\mathcal{L}\to\mathcal{R},
\end{split}
\end{equation}
where $\beta=-\frac{i\eta a}{2\pi}$ and $\mathcal{L}\to\mathcal{R}$ represents the right counterpart where we should make a replacement $(1-e^{-\eta a})\to(e^{-\eta a}-e^{-\eta D})$ and replacement $a\to(D-a)$ for other $a$'s. $\Im$ is the symbol for imaginary part of an expression.

If we make a transform $\eta\to -ib\frac{\pi}D, \Delta\to i\Delta$ and make use of the real part of Eqs.~\ref{eq:fourierexponent} and \ref{eq:dreII4}, we can have some insights for profiles $f(x)=\cos{(\frac{b\pi}D x+\Delta)}$, where $0 \le \Delta < 2\pi$ and $b>0$. We have
\[
\begin{split}
F_{m,b}^\mathcal{L}
&=\frac 1{4\pi}[\sin{\Delta}-\sin{(2\pi\frac{ba}{2D}+\Delta)}]\\
&\cdot\left[\frac 1{m-ba/2\pi}-\frac 1{m+ba/2\pi}\right],
\end{split}
\]
and 
\[
\begin{split}
\delta\tilde{E}_2
&= \frac{\delta\alpha An_0^2}{2\pi}\frac 1{4\pi}[\sin{\Delta}-\sin{(2\pi\frac{ba}{2D}+\Delta)}]\\
&\cdot \partial_\xi^3 \left[\phi(e^{-\frac{\pi\xi}{an_0}},1,-\frac{ba}{2\pi})-\phi(e^{-\frac{\pi\xi}{an_0}},1,\frac{ba}{2\pi}) \right].
\end{split}
\]
Here we choose $b=2$ and $\Delta=0$. For $0<a<D$ we have $0<ba/2D<1$ so the Lerch function is well defined. Together with another profile $f(x)=1$ and some numeric factors, we can recover the result of Ref.~\cite{Horsley2013} following the procedures above.

Back to the exponentially decaying profile, we evaluate the influence of inhomogeneity on the total Casimir force. We focus on the left cavity part. When $a\to\infty$, $\Re[\log{i\beta}-\psi(\beta)]$ vanishes, but $\left(\psi(1)-\log{\frac \eta{2n_0}}+\frac{11}6\right)$ is nonzero and depends on material’s properties $n_0,\eta$. This means Eq.~\ref{eq:dreIIex} contains a part of energy that is not free and thus has no influence to Casimir force. We let $D\to\infty$ to remove the right cavity so that we can focus only on the left cavity (two-plate interaction). The force contributions are
\begin{gather}
F_h=-\frac{A\hbar c \pi^2}{240n_0}\frac 1{a^4},\label{eq:fh}\\
\delta F_1=\delta\alpha\frac{A\hbar c \pi^2}{240n_0}\frac 1{a^4}
\left[F_0^\mathcal{L}+\frac 13(F_0^\mathcal{L}-f(a))\right],\label{eq:f1}\\
\begin{split}
\delta F_2
&=\delta\alpha\frac{A\hbar c \pi^2}{240n_0}\frac 1{a^4}\Bigg\{\frac{60}{\pi}(e^{-\eta a}-1)\Im{\left[\beta^3(1-\beta\psi'(\beta))+\frac{\beta}6\right]}\\
&-\frac{60}{\pi}\eta a e^{-\eta a}\Im\Bigg[\beta^3\left(\psi(1)-\log{\frac \eta{2n_0}}+\frac{11}6\right)\\
&+\beta^3(\log{i\beta}-\psi(\beta))-\frac{\beta}{12}\Bigg]\Bigg\}.\label{eq:f2}
\end{split}
\end{gather}
Eq.~\ref{eq:fh} is the well-known Casimir force between two plates within homogeneous media. If we treat the medium between two plates as homogeneous and use the average refraction index, we can get an approximation of the Casimir force
\begin{equation}
\bar{F}_h=-\frac{A\hbar c \pi^2}{240n_0}\frac 1{a^4}[1-\delta\alpha F_0^\mathcal{L}].
\end{equation}
This is exactly the combination of Eq.~\ref{eq:fh} and the first part of Eq.~\ref{eq:f1}. The rest (second part) of Eq.~\ref{eq:f1} together with Eq.~\ref{eq:f2} is written as $\delta\bar{F}$. This term is easy to understand. It reflects the change of the average refraction index when the plate is shifted.

\begin{figure}[t]
\scalebox{1}{\includegraphics{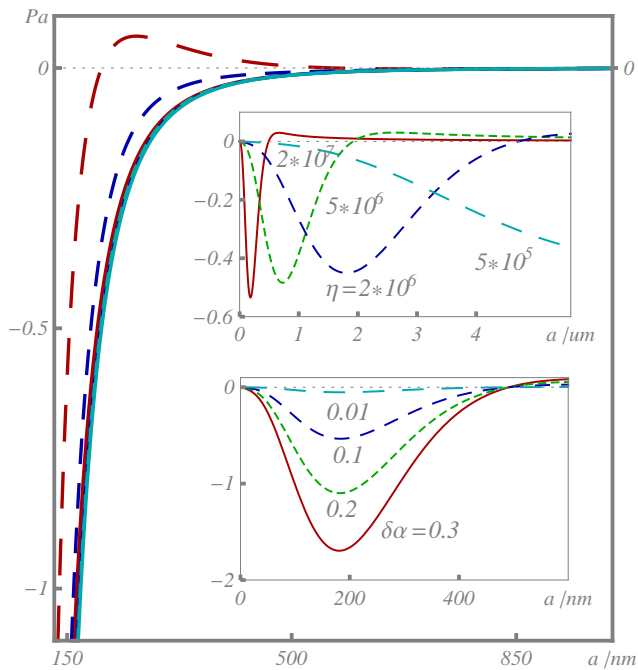}}
\caption{\label{fig:correction}The homogeneous approximation and the exact Casimir pressure including the first-order correction between plates within an inhomogeneous medium. Solid curves are approximations and dashed curves are exact pressures in unit of Pascal (minus means attractive). All $x$ axis is the position $a$. The parameters are $\eta=10^7 /m, n_0=1.5$ for all, while $\delta\alpha=0.3, 0.1, 0.01$ for red (long-dashed), blue (medium-dashed) and cyan (short-dashed) curves respectively. Red solid, blue solid, cyan solid and cyan short-dashed curves appear to coincide. Insets are ratios of $\delta\bar{F}/\bar{F}_h$. Top inset: $\delta\alpha=0.1, n_0=1.5$ and $\eta=2\times 10^7, 5\times10^6, 2\times10^6, 5\times10^5 /m$ for red solid, green short-dashed, blue medium-dashed, cyan long-dashed curves, respectively. Bottom inset: $\eta=2\times10^7 /m, n_0=1.5$ and $\delta\alpha=0.3, 0.2, 0.1, 0.01$ for red solid, green short-dashed, blue medium-dashed, cyan long-dashed curves, respectively.}
\end{figure}

The relation between $\bar{F}_h$ and $\delta\bar{F}$ is given in Fig.~\ref{fig:correction}.
We should emphasize that, according to Eqs.~\ref{eq:polarization} and \ref{eq:fourierexponent}, we have $\delta\alpha\sum_\lambda{\mathcal{P}_J}<2\delta\alpha$. All $\delta\alpha<1$ is permitted within perturbation theory (though the first-order correction might not be enough). Our simulation results clearly show, when $\eta>10^7, \delta\alpha>0.2$, the correction even dominates over the homogeneous approximation and flips the sign in some range. When $\delta\alpha=0.1$, the first-order corrections also have a relative magnitude of peak $50\%$, thus can not be ignored. We should take seriously this inhomogeneity-induced repulsion, since this might indicate alternatively the first-order perturbation is too rough for such an intense inhomogeneity. To investigate whether inhomogeneity can induce repulsive Casimir force and be used to control the Casimir force, we need further studies. On the other hand, when $\delta\alpha\ll 0.01$, or $\eta\ll 5\times 10^5/m$, in the range of 0--1$um$ where the Casimir force is measurable, the correction is well below $1\%$ and can be omitted. Actually, in our minds, most experiments in natural condition are belong to this case.
\section{\label{sec:discussion}discussions}
First, we note the linearly inhomogeneous case $f(x)=x/D$. From Eq.~\ref{eq:dreII3} we know $\delta\tilde{E}_2=0$, thus there is no logarithm term. This seems to tell us that, removing the ``bulk contribution" is enough in this case to retrieve finite Casimir force in the half-space model. This actually is not true. A second-order correction $\langle\chi^0_J|(x/D)^2|\chi^0_J\rangle$ would immediately introduce in logarithm terms, let alone higher order corrections.

We thus stress that our result is valid within the first-order perturbation theory, because, without a tedious calculation of higher order perturbation or any other rigorous demonstration, one can never claim there wouldn't be any omitted term that is actually both $a$-dependent and cutoff-dependent. One example is, $\delta \tilde{E} \sim \xi^{-4}+a\log\xi+f(a)$. When $\xi\to 0^+$, to calculate perturbative energy for the first-order perturbation, $a\log\xi$ may be possibly ignored as a trivial term. Since $\xi^{-4}$ is independent of $a$, it will disappear in the expression of Casimir force. Then everything looks fine and no cutoff dependence is found in such an approach. However, what matters the most to the Casimir force is the derivative of energy with respect to $a$, and if we know the exact perturbative energy somehow, then we see the previously trivial term $a\log\xi$ now arises to be cutoff-dependent.

At last, if somehow we know higher order corrections would not produce both $a$ and cutoff dependent terms, which is a reasonable expect, the magnitudes of first-order corrections would be quite reliable then.
\section{\label{sec:conclusion}conclusions}
With mode-summation technique and first-order perturbation theory, we have expressed the regularized Casimir energy for inhomogeneous case with cylinder kernel coefficients, as Eq.~\ref{eq:HKE}. Like other unresolved Casimir divergences, we found the presence of the logarithmically cutoff-dependent term (see Eq.~\ref{eq:log} and the subsequent analysis). Our results also show there is a term of quadratic cutoff dependence in the Casimir energy. In the piston model such terms are independent of position $a$ and thus vanish in the force, while in the half-space model such terms are dependent of $a$ and thus remain in the force. Consequently, we must introduce additional regularizations to remove them in the half-space model, though, it is not clear how to do it.

Based on the piston model, our results have shown, for any smoothly inhomogeneous profile, the Casimir force is always cutoff-independent in the first-order perturbation. For some other profiles that are not smooth, it seems one can still get cutoff-independent result, though we can not give a rigorous proof that it's always the case yet. Our result supports the method in Ref.~\cite{Goto2012} to omit diverging terms when simulating the Casimir force within inhomogeneous media numerically.

We have also calculated the first-order corrections to both free Casimir energy and Casimir force for exponentially decaying profile. Surprisingly, comparing with the homogeneous analogue where the average refraction index between two plates is used, we found the correction to Casimir force can be even larger than the predicted value of the homogeneous analogue, and flips the sign of the force, though we note the first-order correction might not be accurate enough. All of these results may be useful as a reference for future relevant theoretical calculations and experimental studies.
\begin{acknowledgments}
We wish to thank the program of Zhejiang Leading Team of Science and Technology Innovation.
\end{acknowledgments}

\appendix


\bibliography{Casimir}
\end{document}